\documentclass[12pt,preprint]{aastex}

\begin{document}

\title{Resolved Galaxies in the Hubble Ultra Deep Field: Star
Formation in Disks at High Redshift}

\author{Debra Meloy Elmegreen \affil{Vassar College, Dept. of
Physics \& Astronomy, Box 745, Poughkeepsie, NY 12604;
elmegreen@vassar.edu}}
\author{Bruce G. Elmegreen\affil{IBM Research Division, T.J. Watson
Research Center, P.O. Box 218, Yorktown Heights, NY 10598,
bge@watson.ibm.com} }

\author{Swara Ravindranath
\affil{Space Telescope Science Institute, 3700 San Martin Dr.,
Baltimore,  MD 21218; Inter-University Centre for Astronomy and
Astrophysics; Pune University Campus, Pune, Maharashtra, India
411007, swara@iucaa.ernet.in} }

\author{Daniel A. Coe \affil{Johns
Hopkins University, Dept. of Physics \& Astronomy, 3400 N. Charles
St., Baltimore, MD 21218, USA; Instituto de Astrof\'isica de
Andaluc\'ia (CSIC), C/ Camino Bajo de Hu\'etor 50, Granada 18008,
Spain; coe@iaa.es} }

\begin{abstract}
The photometric redshift distributions, spectral types, S\'ersic
indices, and sizes of all resolved galaxies in the Hubble Space
Telescope Ultra Deep Field (UDF) are studied in order to
understand the environment and nature of star formation in the
early Universe. Clumpy disk galaxies that are bright at short
wavelengths (restframe $<5000$\AA) dominate the UDF out to
$z\sim5.5$. Their uniformity in $V/V_{max}$ and co-moving volume
density suggest they go even further, spanning a total time more
than an order of magnitude larger than their instantaneous star
formation times. They precede as well as accompany the formation
epoch of distant red galaxies and extreme red objects. Those
preceding could be the pre-merger objects that combined to make
red spheroidal types at $z\sim2-3$. Clumpy disks that do not
undergo mergers are likely to evolve into spirals. The morphology
of clumpy disks, the size and separation of the clumps, and the
prevalence of this type of structure in the early Universe
suggests that most star formation occurs by self-gravitational
collapse of disk gas.
\end{abstract}

\keywords{galaxies: formation --- galaxies: evolution ---
galaxies: high-redshift}

\section{Introduction}

Star formation in galaxies today occurs in disks where molecular
clouds contain the densest gas and random motions are much slower
than the rotation speed. In contrast, monolithic collapse models
(Eggen, Lynden-Bell \& Sandage 1962; Larson 1975) for the
formation of galaxies assume that stars form in three-dimensional
geometries which flatten over time as the gas cools. The formation
of stars in molecular clouds that occupy the full height of a
spheroid has not been directly observed, though, even at high
redshifts. Spheroidal systems do form stars (e.g. Ravindranath et
al. 2006; Teplitz et al. 2006), but the youngest stars and the
clouds that form them could be in disks and scattered pieces of
disks that were formerly inside flatter galaxies that merged.
Ellipticals with blue clumps (Menanteau, Abraham, \& Ellis 2001;
Menanteau et al. 2005; Pasquali et al. 2006) could have either
formed these clumps in gas disks or accreted them as smaller
galaxies with gas disks. For example, UDF 900 is an elliptical
galaxy that reveals nuclear spiral arms in what appears to be a
disk in unsharp masked images (Elmegreen, Elmegreen, \& Ferguson
2005). Blue ellipticals out to $z\sim1.5$ (Franceschini et al.
2006) could be aging merger remnants and not three-dimensional
star formation. When star formation sites are observed either
locally or at high redshift, they usually lie in the disks of
galaxies (Driver et al. 2006; Genzel et al. 2003). This is also
the case for both isolated and merging galaxies that are close
enough to have their disks resolved (e.g., the Antennae). Old
globular clusters could have formed in disks too (Kravtsov \&
Gnedin 2005), like their modern-day counterparts. The spheroids
themselves apparently formed during mergers by the scattering of
disk stars (Toomre 1977; Barnes \& Hernquist 1992; Schweizer
1996); they also grew during mergers of other spheroids (e.g.,
Khochfar \& Burkert 2003; Bundy et al. 2004; van Dokkum 2005; Bell
et al. 2006; Robertson et al. 2006; Trujillo et al. 2006b).

Here we are interested in the structure and redshift distribution
of star-forming galaxies that are observed in the Hubble Ultra
Deep Field (UDF). We consider all the resolvable galaxies in the
UDF and compare their photometric redshift distribution with the
sequence of galaxy assembly emerging from recent observations. In this
sequence, giant elliptical galaxies assembled at $z\sim 0.8-2$
from smaller progenitors whose stars formed at $z\sim 1.5-5$ (van
Dokkum et al. 2003; Daddi et al. 2004a, 2005; Fontana et al. 2004;
Stanford et al. 2004; Bundy et al. 2005; Labb\'e et al. 2005;
Saracco et al. 2005; Longhetti, et al. 2005; Cimatti et al 2006;
De Lucia \& Blaizot 2006; Franceschini et al. 2006; Fritz von
Alvensleben \& Bicker 2006; Homeier et al. 2006; Huang et al.
2005; Rigopoulou et al. 2006; Roche et al. 2006; Rudnick et al.
2006). Slightly earlier assembly and star formation occurred in
denser clusters (e.g., di Serego Alighieri, Lanzoni, \& Jørgensen
2006; Clemens et al. 2006; Sheth et al. 2006) and for more massive
galaxies (e.g., van der Wel et al. 2004; Treu et al. 2005; Feulner
et al. 2005; Bundy et al. 2006, Kajisawa \& Yamada 2006;
Franceschini et al. 2006), but even most low-mass ellipticals appear
to have assembled before $z\sim0.8$ (Andreon 2006).

Also in the observed sequence, galaxies with regular disks appear mostly at $z \lesssim 2.5$
(Dawson et al. 2003; Cassata et al. 2005; Conselice et al. 2005a,
2006; Genzel et al. 2006; Ravindranath et al. 2006; Trujillo et
al. 2006a). Some regular disk galaxies are old even at $z\sim
1.5-2.5$ (Iye et al. 2003; Stockton, Canalizo, \& Maihara 2004;
Fu, Stockton, \& Liu 2005; Stockton, McGrath, \& Canalizo 2006;
Toft et al. 2005), indicating their stars formed at $z\sim 5$
(Roche et al. 2006). Sub-mm galaxies at $z\sim2-3$ could also be
disks with active star formation (Almaini et al. 2006; Tacconi et
al. 2006; Swinbank et al. 2006). Peculiar galaxies dominate galaxy
morphologies at $z> 1.5 $ (Abraham et al. 1996a,b; Conselice,
Blackburne, \& Papovich 2005a; Conselice et al. 2005a). Peculiar
and compact types also dominate star formation at $z>1$ (Menanteau
et al. 2006; de Mello 2006). When the peculiar types are resolved,
they are usually clumpy and distorted disks (Elmegreen, et al.
2005a; hereafter EERS).

These redshift distributions for ellipticals, spirals, and
peculiars are consistent with the redshift distributions of major
and minor mergers. According to Conselice et al. (2003), Treu et
al. (2005) and others, the epoch of major mergers ended around
$z\sim2$ and produced most of the giant ellipticals. After this
time, mergers continued to form smaller ellipticals (e.g., Bundy
et al. 2005) and the relatively isolated spirals survived, growing
primarily by dark matter plus gas accretion and minor mergers
(e.g., Conselice et al. 2005b; Brook et al. 2006). Recent
numerical studies indicate a general agreement between the epochs
of elliptical galaxy formation, nuclear black hole activity, and
major mergers (Hopkins et al. 2006).

A key component in this standard model is the physical nature of
star formation in young galaxies.  One way to understand this star
formation is by studying its morphology in the context of the overall galaxy. Galaxies at high
redshift can have morphologies and spectral properties that are
different from those of local galaxies, as shown by recent studies
of the Hubble Deep Fields (HDF-N, Williams et al. 1996; HDF-S,
Volonteri et al. 2000), the Ultra Deep Field (UDF; Beckwith et al.
2006), GOODS (Giavalisco et al. 2004), and other surveys. Some of
this difference results from bandshifting of the UV emission from hotspots into
the optical range (Toft et al. 2005; Cassata et al. 2005), some
from a greater clumpy structure at high redshift, and some from
intrinsically different galaxy shapes. The high resolution of the
Advanced Camera for Surveys (ACS) has enabled morphological
studies of galaxies in the UDF out to a redshift of z$\sim$5,
beyond the primary assembly epoch of most modern spirals and giant
ellipticals.  The physical resolution of the UDF corresponds to
$\sim200$ pc per pixel ($0.03^{\prime\prime}$) for $0.5<z<5.5$, so
we expect to resolve the disks and see directly the star formation
properties of the most common progenitor galaxies.

There are many methods for studying the overall morphology of
galaxies, but comparatively few for the morphology of star
forming regions within them. For overall morphology, visual inspection provided the
initial basis for the main galaxy classifications (Sandage 2005;
de Vaucouleurs et al. 1991; van den Bergh 1960) and it is still
the most useful method if there are highly resolved structures
such as rings, bars, and spiral arms. High redshift galaxies
typically have less resolved structure and they also include
peculiar clumpy types that are not found locally (Abraham et al.
1996a,b; Cowie, Hu, \& Songaila 1995; van den Bergh 2002; van den
Bergh et al. 1996). Thus the Hubble class and other standard
systems have limited value at high redshift.

New systems have been developed for galaxy classification at high
redshift that rely on automated measures of certain simple
properties. These properties include concentration, asymmetry, and
clumpiness in the CAS system (Conselice 2003, Conselice et al.
2004 and references therein; Pasquali et al. 2006), the S\'ersic
index for the surface brightness profile (S\'ersic 1968; see,
e.g., Trujillo et al. 2004; Ravindranath et al. 2006), the
restframe spectral energy distribution (Ben\'itez 2000; Coe et al.
2006), and the Gini coefficient, which measures the relative distribution of
brightness from pixel to pixel (Abraham et al. 2003; Lotz et al.
2006). These systems are interrelated. Conselice (2003), Menanteau
et al. (2006) and others show there is a correlation between the
asymmetry and concentration value of a galaxy and the Hubble
classification as a spiral, elliptical/S0, or Irr/peculiar. However
Franceschini et al. [2006] found
that 30\% of $z>1.3$ galaxies classified as spheroids by CAS look
more like late Hubble types visually, although only 5\% of the
reverse was true.
S\'ersic indices, $n$, also correlate with Hubble type, but
$n$ has a wide range of values
among high redshift elliptical and spiral galaxies (e.g.,
Hatziminaoglou et al. 2005) unlike local galaxies where $n\sim4$
generally corresponds to an elliptical and $n\sim1$
is typically seen for a spiral.

The star formation morphologies of galaxies are not included in
these classification systems. One has to rely instead on visual
inspection of images, aided by color maps and contrast enhancement
techniques. These images suggest that star-forming regions in most
high redshift galaxies are larger and more massive than in
galaxies today (see review in Elmegreen 2007). Star-forming
galaxies are also disk-like, ranging from edge-on clumpy disks
that can probably be identified with chain galaxies (Reshetnikov,
Dettmar, \& Combes 2003; Elmegreen, Elmegreen \& Hirst 2004a) to
face-on clumpy disks that have been
referenced by various names in the literature, including spirals,
protospirals (van den Bergh 2002), luminous diffuse objects
(Conselice et al. 2004), and clump clusters (Elmegreen, Elmegreen,
\& Sheets 2004). Clump clusters contain no bulges or exponential
light profiles, unlike spirals at the same redshifts, and
$\sim25$\% of their $i_{775}$ light is in clumps, compared to
$\sim 5$\% for spirals (Elmegreen et al. 2005b).

The clumps in high redshift galaxies are usually $\sim1-2$ kpc in
diameter in the restframe UV images, and they are bright with star
formation ages of $\sim300$ My and masses of $\sim10^8-10^9$
M$_\odot$ (Elmegreen \& Elmegreen 2005; hereafter EE05). A typical
clump cluster galaxy may contain only $\sim5$ of these giant
clumps at any one time, making its surface brightness extremely
irregular, yet the mean radial profile of the clumps is an
exponential like the smooth profiles in high redshift spirals
(Elmegreen et al. 2005b). Clump sizes at high redshift are also
about the same as edge-on galaxy thicknesses, $\sim1$ kpc for a
sech$^2$ scale height, and the clump centers are highly confined
to the midplanes (Elmegreen \& Elmegreen 2006; hereafter EE06).
These morphologies make the clumps look like they continuously
form in a thick disk gas layer, probably by gravitational
instabilities considering their length and mass scales (EE05; EE06). As they
dissolve they build up the stellar exponential disks. For example,
some giant clumps in clump cluster galaxies have head-tail
structures that suggest tidal forces are hastening their
dissolution (EE05).

In this paper, we are interested in the redshift distribution of
clumpy star-forming galaxies in the UDF. Similar studies for
spirals and ellipticals in various HST deep fields have been
available for several years (e.g., Roche et al. 1998; van den
Bergh 2002; Kajisawa \& Yamada 2001; Conselice, Blackburne, \&
Papovich 2005a; Stanford et al. 2004; Cassata et al. 2005;
Franceschini et al. 2006). Consistent with these other studies, we
find a wide range of S\'ersic indices and star formation rates for
the spirals and ellipticals in the UDF, indicating a broad range
of structures and formation redshifts, and we find that the more
irregular morphologies dominate at higher redshift.

What is new here is the observation that extremely young disks
extend back to redshifts that are as early as the first main
star formation phases in high mass elliptical galaxies.  This
implies that the elliptical and proto-elliptical galaxies observed
at intermediate redshifts (e.g., the distant red galaxies; Franx
et al. 2003; F\"orster Schreiber et al. 2004) could have formed by
mergers of clumpy disk systems like those observed here out to
$z\sim5$. A $V/V_{max}$ test suggests we have not even reached the
surface brightness limit of these clumpy disks, but only the bandshifting limit
from the redshift of restframe UV light out of the ACS
optical bands.
These observations reinforce the idea that most
star formation occurs in disks and
that clumpy disks may host even the earliest Population II stars.

In what follows, the data are presented in Section 2, the results
and analysis are in Section 3, and a discussion of the
implications is in Section 4. We assume a $\Lambda$CDM model with
$\Omega_\Lambda=0.73$, $\Omega_M=0.27$, and $H_0=71$ km s$^{-1}$
Mpc$^{-1}$ (Spergel et al. 2003).

\section{Data and Analysis}

The Hubble UDF images were used for this study
(Beckwith et al. 2006).  The images
consist of exposures in 4 filters:  F435W (hereafter $B_{435}$;
134880 s exposure), F606W ($V_{606}$; 135320 s), F775W ($i_{775}$;
347110 s), and F850LP ($z_{850}$; 346620 s). The images have a
size of 10500 x 10500 pixels with a scale of 0.03 arcsec per px
(315 arcsec x 315 arcsec); they were obtained from the STScI
public archive.

\begin{figure}\epsscale{0.8}
\plotone{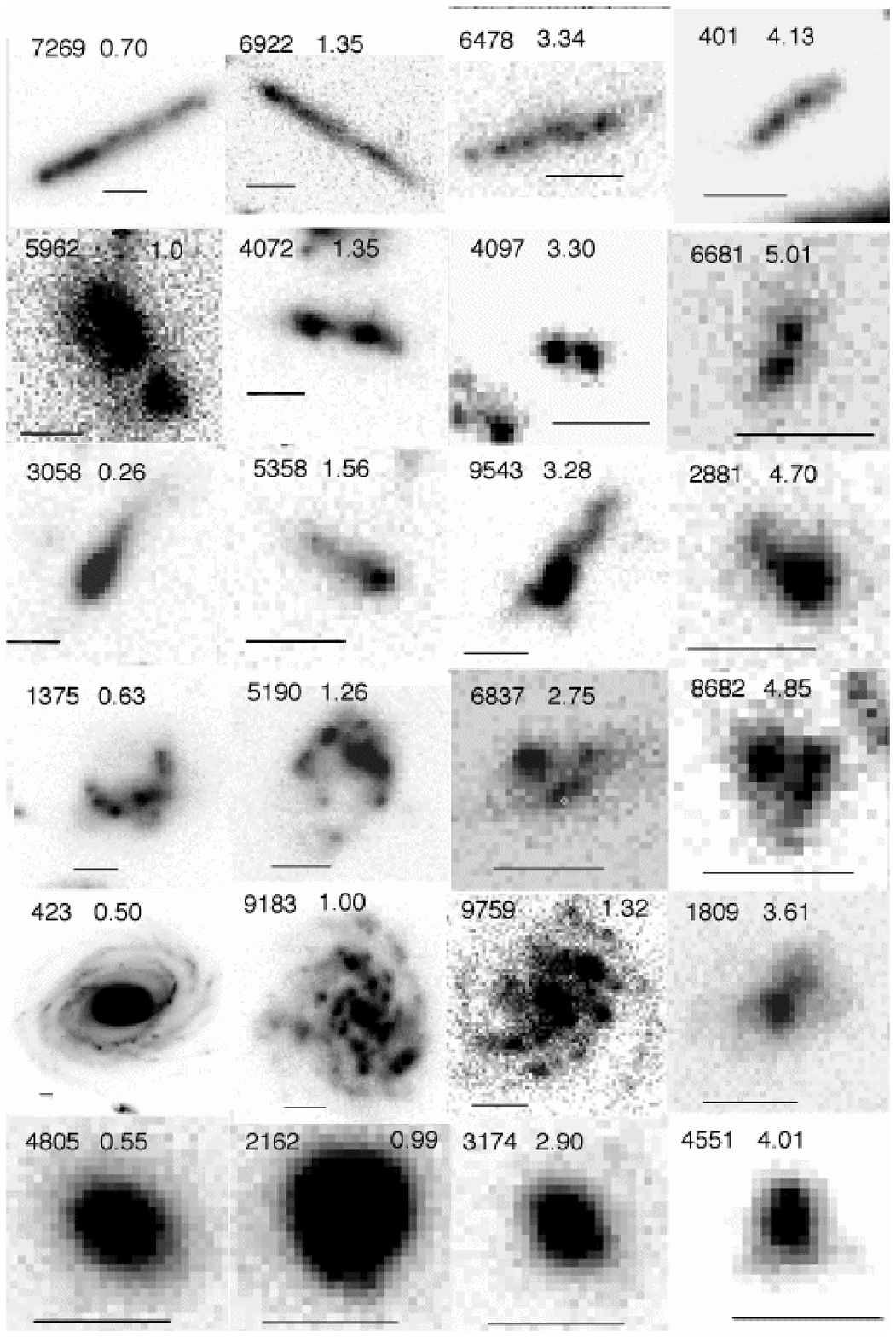} \caption{Examples of each morphological type are
shown from $i_{775}$ images.  From top to bottom, each row
contains: chains, doubles, tadpoles, clump clusters, spirals, and
ellipticals. The UDF catalog number is in the upper left of each
image, along with the redshift, which increases from left to
right. The bar indicates 0.5".}
\label{fig:UDF_zmontage}\end{figure}

In the UDF, all galaxies (884) larger than 10 pixels in diameter
were classified by visual inspection from the $i_{775}$ images
into one of 6 main types: chains, with prominent clumps in a
straight row; doubles, with two prominent, apparently connected
clumps; clump clusters, with several apparently connected clumps
in a 3-D or disk arrangement; tadpoles, with one prominent clump
and a tail of smaller clumps or diffuse emission; spirals, with a
nucleus, spiral arm-like structure and an exponential-like
disk profile; and ellipticals, with a spheroidal central light
concentration based on concentric elliptical light contours. The
resulting catalog (EERS) provides the basis for what will
subsequently be referred to as the morphological class for each
galaxy. For the present paper, we re-examined all the entries for
positional accuracy, removed any entry that had clumps with highly
discrepant photometric redshifts (suggesting line of sight
alignments), and expanded the list to include more galaxies at the
10 px limit. The current revised catalog includes 1003 galaxies
(121 chains, 134 doubles, 192 clump clusters, 114 tadpoles, 313
spirals, and 129 ellipticals). Examples of each type over a range
of photometric redshifts are shown in Figure
\ref{fig:UDF_zmontage} in the $i_{775}$-band. Note that the
highest $z$ spiral does not have a well-defined spiral structure;
it is classified as a spiral because of its central clump and
exponential light distribution.

The primary selection limits are in size (2-$\sigma$ isophotal
contour along the major axis greater than 10 px) and surface
brightness (brighter than $26.0$ mag arcsec$^{-2}$ at the
2-$\sigma$ contour in $i_{775}$ band). On a plot of magnitude
versus size (e.g., Ravindranath et al. 2004), our selection
corresponds approximately to galaxies brighter than a certain
magnitude limit for each size, with brighter apparent magnitude
limits corresponding to larger sizes. This limit has a broad
spread, however, because linear galaxies (chains, tadpoles,
doubles, and edge-on spirals) have fainter magnitudes than
circular galaxies for a given size and surface brightness.

Redshifts were determined using the Bayesian Photometric Redshift
(BPZ) method (Ben\'itez 2000; Coe et al. 2006), based on the ACS and
NICMOS photometric data for the catalogued galaxies. New
segmentation maps were made for most of the galaxies in our
catalog, starting with central coordinates and estimated sizes.
They were needed for the clumpy galaxies, which SExtractor often
subdivided into separate objects. Some of the galaxies listed in
the online UDF catalog consist of more than one object or clump at
the same redshift. Because they share a 2-$\sigma$ intensity envelope,
we consider them to be single galaxies, as in
EERS. The BPZ method yields a probability for the redshift of each
galaxy, and a spectral type ($t_b$) from 1 to 8 that is based on
restframe colors. The error in photometric redshift is not large
enough to affect any of the discussion here; it is estimated to be
$\Delta z\sim0.04\left(1+z\right)$ (Coe et al. 2006).

The S\'ersic index, $n$, classifies galaxies by their light
distributions with intensity $I \propto \exp\left( -r^{1/n}
\right)$. A S\'ersic index of 1 corresponds to an exponential
light distribution while a 4 corresponds to the de Vaucouleurs
(1948) $r^{1/4}$ law for ellipticals. Spheroid-dominated galaxies
are generally considered to be those with $n > 2.5$, while
disk-dominated galaxies normally have $n < 2.5$ (e.g.,
Ravindranath et al. 2006; Aceves, Velazquez, \& Cruz 2006).
S\'ersic fits have been made to a large sample of galaxies in the
GOODS field (Ravindranath et al. 2004, 2006; Cassata et al. 2005).

Galaxies in the present study were fit with the 2-dimensional
routine GALFIT (Peng et al. 2002) to determine a S\'ersic index
and effective radius, $r_{eff}$, defined as the half-light radius
based on the total integrated flux of the best-fit model for the
surface brightness distribution. For galaxies with $z<1.31$, the
fits were done on the restframe B data. This means that for
$z<0.24$, the $B_{435}$-band data were used to make the fits. For
$z$ in the range from 0.24 to 0.56, $V_{606}$-band data were used
because the restframe $B$ band is centered on the observed
$V_{606}$ band for $z=0.4$. For $z$=0.56 to 0.61, an interpolation
between $V_{606}$ and $i_{775}$ was used to give the result in the
restframe $B$. For $z$=0.61 to 0.93, $i_{775}$ band data alone
were used. For the small range between $z=0.93$ to 0.94, an
interpolation between $i_{435}$ and $z_{850}$ was done. Finally,
for $z$=0.94 to 1.31, $z_{850}$ band data was used. Galaxies with
$z>1.31$ do not have restframe B data, so the fits were all done
for $z_{850}$ in those cases. The resulting restframe UV intensity
profile could differ from the restframe B-band profile if there is
a red bulge or other old component in the disk that is not
prominent in the restframe UV (Giavalisco et al. 1996; Kuchinski et al.
2000; Lotz et al. 2006).

UDF simulations by Coe et al. (2006) reveal the accuracy to which
the S\'ersic index $n$ can be measured. They find $n$ is retrieved
to within $\pm1$ for bright and large galaxies ($i_{775}<26$ mag
arcsec$^{-2}$; $r_{eff}>10$ pixels) and to within $\pm2$ for faint
and small galaxies ($i_{775}>28$ mag arcsec$^{-2}$; $r_{eff}<3$
pixels). For UDF galaxies with relative uncertainties
$\Delta(n)/n<1$, 80\%-95\% will be correctly classified as late
($n<2.5$) or early ($n>2.5$). Here the percentage should be
towards the high end of this range because we include mostly
bright UDF galaxies.

\section{Results}
In the following sections, we present and discuss the properties
of UDF galaxies that are derived from BPZ redshifts and GALFIT
profile fitting.  These properties reveal that clumpy disk
galaxies, the chains and clump clusters, extend to $z\sim5$ as
strong starburst systems, going far
beyond the optically observable range for most spirals and
ellipticals, and also farther than the epoch of major mergers at
$z\sim1-3$ when present-day large galaxies were assembled.  The
implications are that the most primitive galaxies are small disks,
and the earliest star formation occurs in disks. The results are
discussed in Sect. \ref{sect:disc}.

\subsection{Redshift Distributions}
\begin{figure} \epsscale{0.8}
\plotone{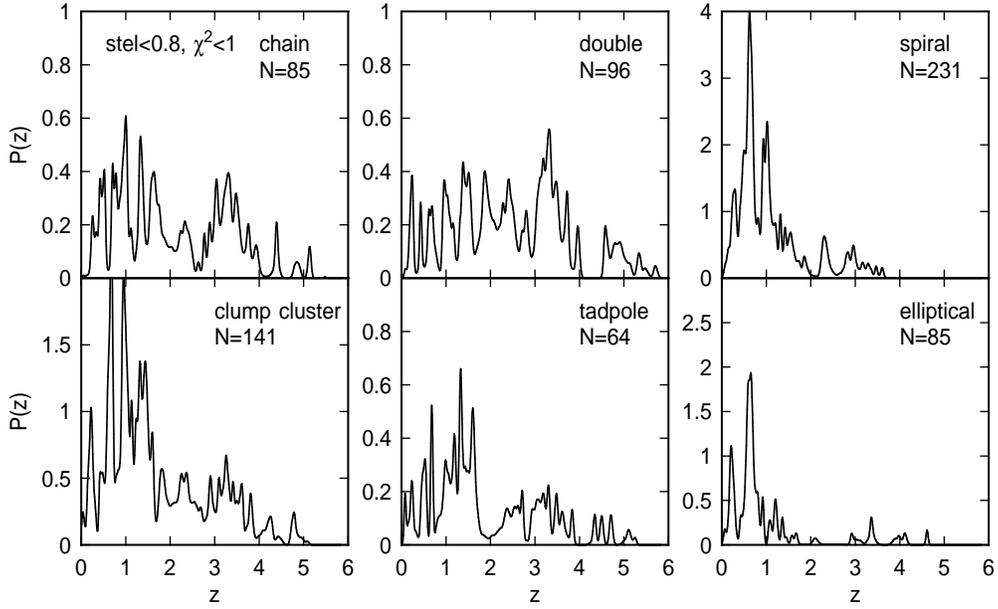} \caption{Probabilities for redshift $z$ are shown
for the 6 morphological types. BPZ returns an entire probability
distribution $P(z)$ for each galaxy.  The sum of these $P(z)$ are
shown here for each morphological type. Most galaxies have only a
narrow allowed range of $z$, so the distribution of most-probable
redshift looks the same as this distribution of probabilities. The
resolution in $z$ is 0.01. Only galaxies with $\chi_{mod}^2$
(goodness of fit) $<1$ and $stel<0.8$ are plotted here and in the
other figures. The number of galaxies of each type is indicated.
The irregular types extend to larger redshifts than the
ellipticals and spirals in our sample.}
\label{fig:UDF_danpz}\end{figure}

Figure \ref{fig:UDF_danpz} shows the probability distribution
function (pdf) for photometric redshift in each morphological
type. Each pdf is constructed from the sum of the pdfs for the
galaxies of that type that have the most reliable BPZ fits.  Here
and throughout this paper, we consider only these most reliable
BPZ fits, which have $\chi_{mod}^2 < 1$ and SExtractor stellarity
$stel<0.8$. The resolution in $z$ is 0.01 and the redshift error,
as noted above, $0.04\left(1+z\right)$ (Coe et al. 2006).  The
distribution of the most-probable redshift (one value per galaxy)
looks virtually identical to the plotted distribution of redshift
probabilities.

The average redshift for the spirals is about 1, although there
are a few at higher $z$. This local bias is partly the result of a
selection effect against face-on spirals at large $z$ resulting
from surface brightness dimming (EERS). Small galaxies are also
missing because our catalog has a minimum diameter of 10 pixels,
which corresponds to 2.6 kpc at $z=1.5$. Bandshifting causes
spirals to drop out of the $z_{850}$ band beyond $z\sim1$ if their
disks are not intrinsically bright in the UV. Thus Figure
\ref{fig:UDF_danpz} indicates that large UV-bright spirals are
relatively rare at $z>1.5$, compared to the more irregular types,
which are UV-bright. The redshift distribution for ellipticals is
about the same as for spirals.

The linear and clumpy morphologies (chain, clump cluster, tadpole and double)
can be observed to $z\sim5$.
These four types dominate at faint magnitudes too, out to 28 AB
mag on the $i_{775}$ images (EERS). The large redshift range
suggests they either form at high redshift and evolve slowly or
they form continuously over a range of redshifts. Star formation
rates estimated previously (EERS) and in other studies (e.g.,
Daddi et al. 2004a,b; Greve et al. 2005) suggest that galaxies
with intermediate to high masses end their star formation
relatively quickly, in $\sim2$ Gyr or less (e.g., Feulner et al.
2005).  In that case the disks we are observing out to $z\sim5$
probably form or light up over a wide range of redshift. This is
consistent with observations by Reddy et al. (2006a) who found a
wide range of star formation ages for galaxies of various sizes at
$z\sim2$, with van Dokkum et al. (2006), who found the same for
massive galaxies, and with Shapley et al. (2005) who considered
episodic star formation in $z\sim2$ galaxies.

\begin{figure} \epsscale{0.8}
\plotone{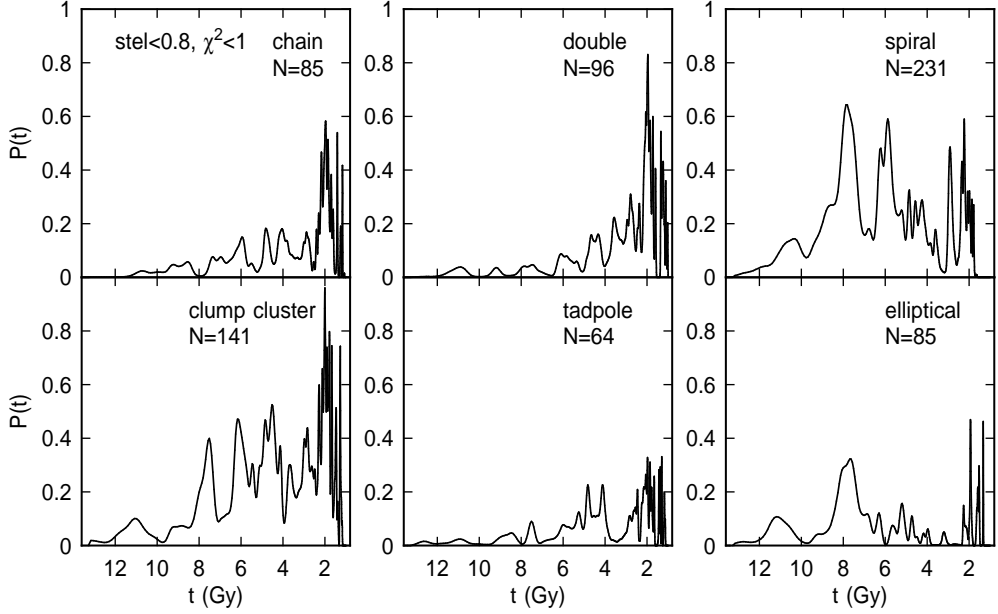} \caption{Probabilities for galaxy age $t$ (since
the beginning of the Universe at $t=0$) are shown for the 6
morphological types. } \label{fig:UDF_danpt}\end{figure}

Figure \ref{fig:UDF_danpt} shows the same distribution functions
as in Figure \ref{fig:UDF_danpz} but plotted versus the age of the
Universe at the redshift of the galaxy, with $t=0$ for the
beginning of the Universe. The values are equal to those plotted
previously, but multiplied by $dz/dt$ for $t$ in Gyr and $z(t)$
the redshift versus age function from the $\Lambda$CMD cosmology.
The linear and clumpy morphologies form quickly and remain common
in the Universe for 6 to 8 Gyr. There are apparently no local
galaxies like chains or clump clusters (e.g., Smith et al. 2001).

\subsection{Magnitude Distributions with Redshift}
\label{sect:mag}

\begin{figure}\epsscale{0.8}
\plotone{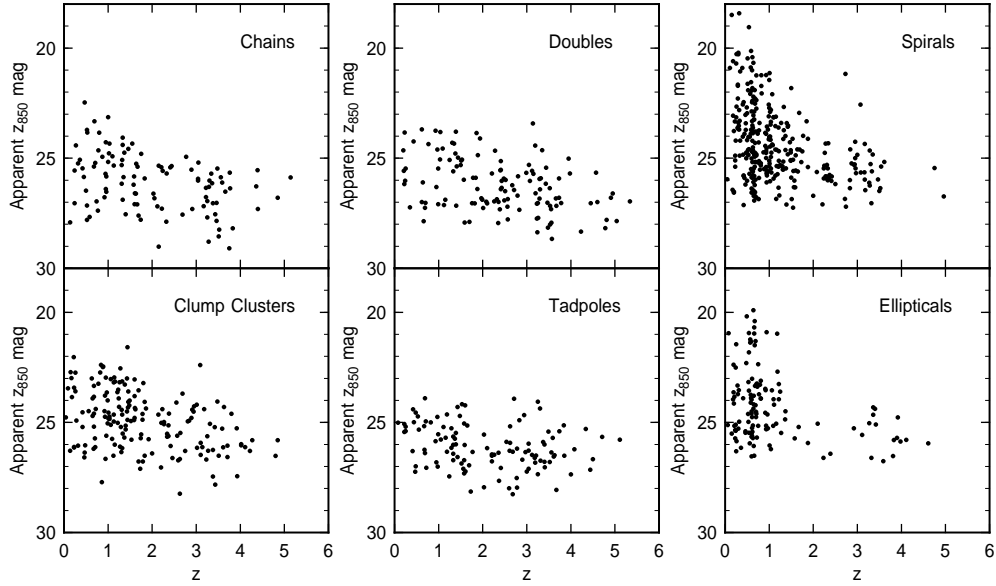} \caption{Redshift distributions of apparent
$z_{850}$ magnitudes for each morphological class. The brightest
ellipticals and spirals are brighter than the brightest clumpy
galaxies at low redshift. At high redshift, all the galaxy types
have about the same magnitudes, although the spirals, ellipticals
and clump clusters are still slightly brighter than the other
three. The distribution of absolute magnitude is shown in Fig. 7.}
\label{fig:danmag}\end{figure}

Figure \ref{fig:danmag} shows the apparent $z_{850}$ magnitude
distribution as a function of redshift for our sample, sorted by
morphological type.  The brightest ellipticals and spirals with
$z_{850}=20$ to 22 mag tend to be brighter than the brightest
clumpy galaxies at low redshift, by 2 to 4 mag. At high redshift,
all the galaxy types have about the same magnitude, 25 to 27 mag,
although the spirals, ellipticals and clump clusters are still
slightly brighter than the other three types, by about 1 mag on
average. Presumably clump clusters are brighter than chains
because the clump clusters are face-on versions of chains (see
radiative transfer models for these inclination effects in
Elmegreen, Elmegreen, \& Hirst 2004a).

Adelberger et al. (2005) studied the clustering properties of
optically selected star-forming galaxies at $z=1.4-3.5$ and found
that they are likely progenitors of intermediate-mass elliptical
galaxies by the time $z=1$.  Their sample consisted of galaxies
with apparent R-band magnitudes of 23 to 25. For Bruzual \&
Charlot (2003) rest-frame population colors $0.17\mu-0.22\mu\sim
0.2$ at 0.1 Gyr and $\sim0.6$ at 1 Gyr, which is the appropriate
$R-z_{850}$ color conversion for $z\sim2$, the Adelberger et al.
sample in R-band is significantly brighter than our
color-corrected sample in R-band for all but the brightest low
redshift spirals and irregulars. This suggests that the clumpy
types in our survey are not as massive as the galaxies in the Adelberger
et al. survey and should not have the same strong clustering
properties.

\subsection{Spectral Type Distributions with Redshift}
\label{sect:tb}

\begin{figure}\epsscale{0.8}
\plotone{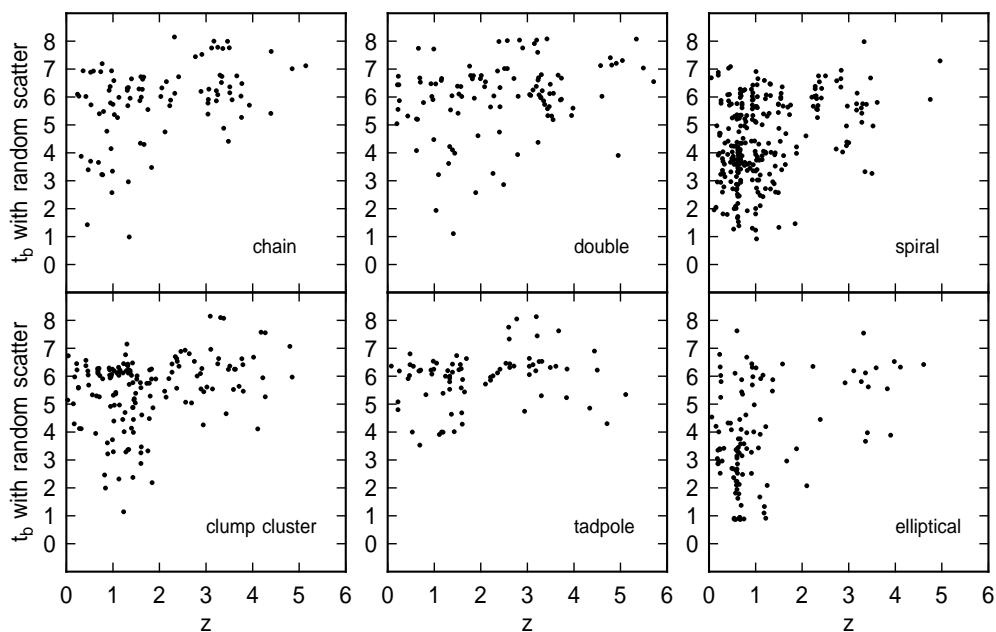} \caption{Distribution of spectral type $t_b$ as a
function of redshift for different morphological types (a random
scatter of $\pm0.15$ has been added to $t_b$ to avoid overlaps).
The reddest $t_b$ types, 1, 2, and 3, only are present at
redshifts less than $z=2.5$, and most of these are spirals and
ellipticals. Starbursts dominate at $z>2$, most likely because of
a bandshifting selection effect.}
\label{fig:UDF_DANTBZ6}\end{figure}

Figure \ref{fig:UDF_DANTBZ6} shows the redshift distribution of
spectral type, $t_b$, for each morphological type. The redshift
for each galaxy is taken to be the peak in the $P(z)$ distribution
used for Figure \ref{fig:UDF_danpz}. The spectral types correspond
to local equivalent types of $1=$elliptical, $2-3=$spiral,
$4=$Irregular-Magellanic, $5-8=$starburst ($5=$SB3, $6=$SB2,
$7=25$ Myr evolved from the burst, $8=5$ Myr evolved; Coe et al.
2006). For clarity in plotting, the spectral type index $t_b$ has
been given a random scatter of $\pm0.15$.

The chain, double, tadpole, and clump cluster morphologies all
show concentrations around $t_b=6$, indicating they are
starbursts. Very few have spectral types like local early-type
spiral and elliptical galaxies. The spiral and elliptical galaxies
span a wider range of $t_b$ than the more clumpy morphologies.
There are many old population types $t_b=1-3$ among the
low-redshift ellipticals, as expected, but there also are
starburst ellipticals at all redshifts. About half of the UDF
ellipticals in this starburst range have blue clumps (Elmegreen,
Elmegreen, \& Ferguson 2005), although other clumpy ellipticals
have smaller $t_b$. The most prominent elliptical galaxy
concentration is at $t_b=3.5$, signifying colors equivalent to
late-type spirals or irregulars. A similar conclusion was made by
Menanteau et al. (2004; 2006), who found
blue irregularities in 30\% to 50\% of the ellipticals in the
Tadpole galaxy field and the parallel NICMOS fields of the UDF,
respectively, up to $z\sim1.3$. Cross et al. (2004) also found
30\% of the E/S0 types at $z=0.5-1$ to be blue in a wide-area deep
ACS survey, and Franceschini et al. (2006) found that 30\% of
spheroidals and bulge-dominated disks at $z\sim1$ in have
relatively blue colors.  The spirals also have a concentration at
$t_b=3.5$ and a broad extension into the starburst range. Only one
spiral is as red as the reddest ellipticals.

Beyond $z\sim2$, most galaxies in our survey are starbursting, but
this is partly the result of band shifting where intrinsically red
galaxies move out of the ACS spectral range. Near-IR surveys show
a population of massive red galaxies at $z>2$ (Labb\'e et al.
2005) that are barely visible in optical surveys (Franx et al.
2003; Rudnick et al. 2006). Kriek et al. (2006) found red galaxies
without emission lines at $z=2-2.9$ and consistent with no star
formation. Still, Reddy et al. (2005) noted a 70\%-80\% overlap in
the identification of star-forming galaxies selected by visible
and near-IR surveys at $z\sim2$, so our ACS survey should not miss
a high fraction of the high-$t_b$ luminous galaxies at that
redshift. Even in the IRAC bands of the Spitzer Space Telescope,
massive spheroidal galaxies begin to decrease in number at
$z\sim1.5$, as do spirals (Franceschini et al. 2006). Daddi et al.
(2005) and Cassata et al. (2005) found the same decrease in a
K-band survey. This decrease is about the same as we see in the
ACS, so most of our drop in elliptical and spiral numbers at $z>2$
is probably from a lack of galaxies and not just bandshifting.

The most massive galaxies at all redshifts tend to have the oldest
stellar populations (Feulner et al. 2005; Drory et al. 2005;
Franceschini et al. 2006; Kajisawa \& Yamada 2006), but there are
still massive galaxies at $z\sim2$ and beyond that have large star
formation rates, on the order of 100 M$_\odot$ yr$^{-1}$ or more
(e.g., van Dokkum et al. 2004; Fontana et al. 2004; Feulner et al.
2005; F\"orster Schreiber et al. 2004; Huang et al. 2005; Kong et
al. 2006; Reddy et al. 2006b; Rigopoulou et al. 2006; Webb et al.
2006).  This is consistent with the presence of high $t_b$
galaxies out to $z\sim5$ in Figure \ref{fig:UDF_DANTBZ6}.

\subsection{Redshift Distribution of Comoving Density}

\begin{figure}\epsscale{0.8}
\plotone{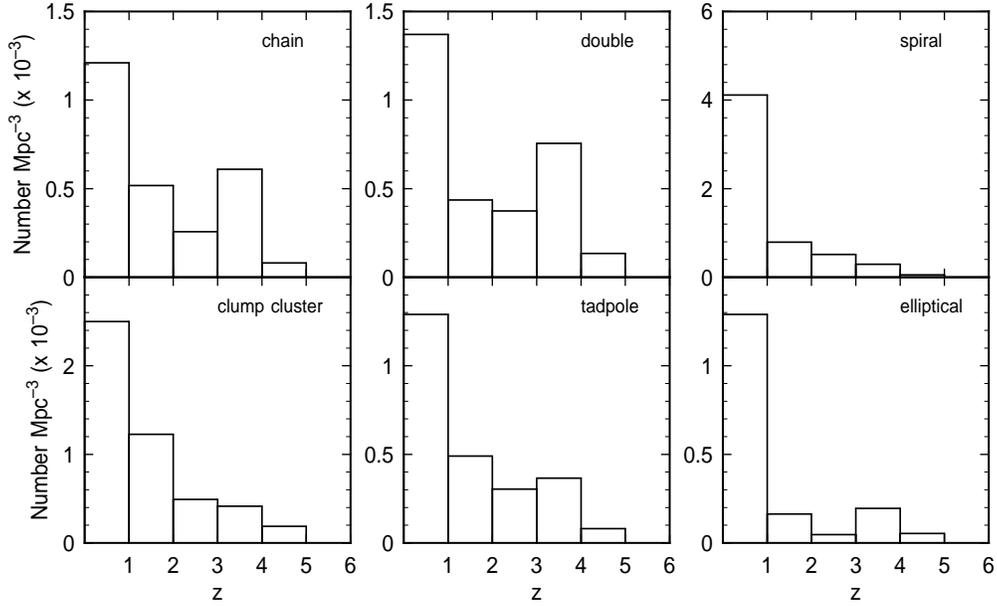} \caption{The comoving volume density of starburst
galaxies ($t_b\geq5$) as a function of redshift. The density
decreases faster with redshift for spirals and ellipticals than it
does for clumpy galaxies.} \label{fig:UDF_DANTBDZ}\end{figure}

The comoving density of each morphological type with starburst
spectra, $t_b\geq5$, is plotted versus redshift in Figure
\ref{fig:UDF_DANTBDZ}. We restrict ourselves to this starburst
range for $t_b$ so that bandshifting out to $z\sim5$ does not
systematically remove galaxies from the ACS images. Ferreras et
al. (2005b) considered comoving density for all early type
galaxies in the UDF and found a decrease from $z=0.6$ to 1.2.

The basic equations for the co-moving volume per unit solid angle
and per unit redshift were obtained from Carroll, Press \& Turner
(1992). We used a solid angle corresponding to 11.97 arcmin$^2$,
which is the part of the UDF that has at least half of the average
depth of the whole image (Coe et al. 2006).  Then the comoving
volume per unit redshift, $dV/dz$, was determined from the basic
equations, and the integrated volume was found from the integral
over $z$: $V(z)=\int_0^z \left(dV/dz\right) dz$. To find the
density, we counted the number of galaxies of each morphological
type with $t_b\geq5$ using equal intervals of $z$ ($0-1$, $1-2$,
etc.), and divided by the comoving volume sampled by this $z$
interval ($V[1]-V[0]$, $V[2]-V[1]$, etc., respectively). The
resulting densities are typically several times $10^{-4}$
Mpc$^{-3}$. Note that the vertical scale in Figure
\ref{fig:UDF_DANTBDZ} is larger for spirals and clump clusters,
which are more abundant than the other types in the UDF.

The comoving densities systematically decrease to higher $z$, but
the decrease is much faster for spirals and ellipticals than it is
for clumpy irregulars. This is not a bandshifting effect because
we have plotted only galaxies that are bright in the restframe uv
($t_b\geq5$). The observed decrease for spirals and ellipticals is
a combination of other selection effects and a real decrease in
these types.

\begin{figure}\epsscale{0.8}
\plotone{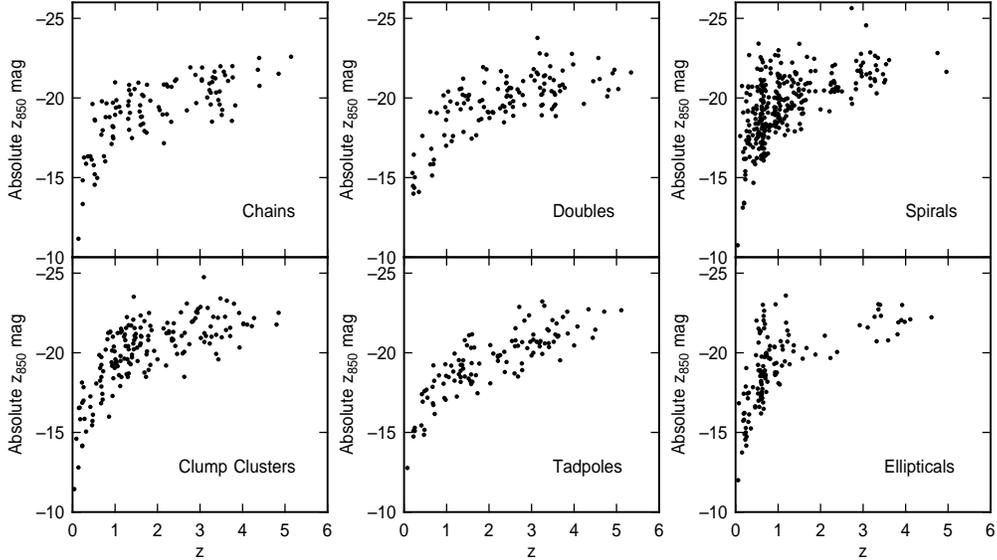} \caption{Redshift distributions of absolute
$z_{850}$ magnitudes for each morphological class. The brightness
increase with $z$ is mostly the result of surface brightness
selection. } \label{fig:danmaga}\end{figure}

Our primary selection effect is the surface brightness limit of
the UDF survey. The angular size limit (10 px) is not severe
because galaxy angular sizes for a given physical size do not
change significantly for $z$ in the range from 1 to 5.  On the other hand,
the observed surface brightness gets faint quickly for a given
intrinsic surface brightness, as $10\log\left(1+z\right)$ in
magnitudes arcsec$^{-2}$.  Figure \ref{fig:danmaga} shows the
absolute magnitudes of all our galaxies, obtained from Figure
\ref{fig:danmag} by applying the distance modulus for the standard
cosmology (Spergel et al. 2003) to the observed $z_{850}$ magnitude
(which is in the restframe UV). The rapid rise in absolute
magnitude at high $z$ indicates that we are selecting only the
brightest galaxies at high $z$, and it also reflects the SED shape
of a young galaxy or starburst. For a starburst SED, the high
redshift emission in the z$_{850}$ band, which corresponds to
the restframe uv, is intrinsically brighter than the low redshift
emission in the z$_{850}$ band, which corresponds to a restframe
visible or red.

Cosmic variance in our sample can be estimated from Somerville et
al. (2004). Considering chain galaxy counts as representative, we
derive cosmic variance fluctuations ranging from $\sim0.6$ counts
at $z=0-1$ to $\sim0.4$ counts at $z\sim4-5$.  These variations
are not large enough to affect the overall trends in Figure
\ref{fig:UDF_DANTBDZ}.

Other selection effects should be considered. Highly clumped
galaxies are easier to see than smooth galaxies when there is
surface brightness dimming because the clumps stand out as local
islands of brightness. Surface brightness dimming makes the
interclump emission fainter while the clumps remain visible,
although at low intensity.  If these clumps are smoothed over into
a region like a spiral galaxy disk with the same average surface
brightness, then this average will fall below the survey limit
before the individual clump brightness does. We noticed this same
effect in the distribution of axial ratios (EERS): face-on spiral
galaxies are relatively rare compared to edge-on spirals in the
UDF, but face-on clumpy galaxies (the clump clusters) have their
expected abundance compared to the edge-on clumpy galaxies (the
chains). The reason for this difference is that face-on spirals
have lower surface brightness than edge-on spirals, so they
disappear from the UDF survey where the a surface brightness limit
is close to the average. The clumps are nearly always visible in
the clumpy galaxies, however, regardless of the orientation of the
surrounding disk. This difference in surface brightness dimming
for clumpy and smooth galaxies could partly explain why the
comoving densities of the four clumpy types with $t_b\geq5$ do not
decrease as fast as the comoving densities of spirals and
ellipticals with $t_b\geq5$.

\begin{figure}\epsscale{0.8}
\plotone{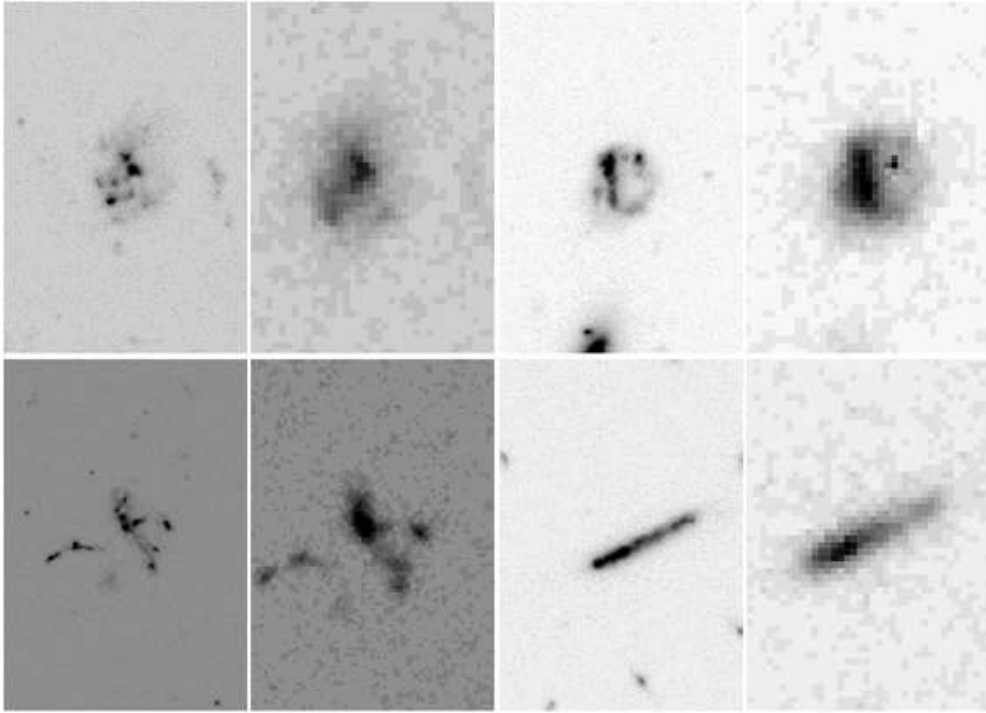} \caption{Three clump clusters and one chain
galaxy are shown with ACS $i_{775}$ on the left of each pair and
NICMOS $J$ on the right. The overall morphology as clump cluster
galaxies does not change significantly at longer wavelength,
suggesting that the relatively rapid drop in comoving spiral
galaxy density is not the result of misclassification. The galaxy
UDF numbers and their photometric redshifts are: upper left, UDF
1666 ($z=1.38$), upper right, UDF 3483 ($z=1.80$), lower left, UDF
6462, ($z=1.43$), lower right, UDF 7269 ($z=0.69$).}
\label{fig:nicmos}\end{figure}

A third effect to consider for Figure \ref{fig:UDF_DANTBDZ} is
that some of the highest-$z$ clump clusters and chain galaxies
could be the star-forming parts of spirals (e.g., Toft et al 2005;
Cassata et al. 2005). The red bulges and underlying red disks of
these spirals could be invisible in the ACS. Such a change in
morphological type with redshift would flatten the comoving
density distribution of the clump clusters and chains, while
steepening it for the spirals. We checked for this possibility by
viewing the NICMOS J-band images of clump cluster galaxies. The
NICMOS images are not high enough resolution to see the details of
star formation (0.09 arcsec px$^{-1}$), but they should resolve
bulges and see the underlying smooth red disks. We found that
clump cluster and chain galaxies look about the same in the NICMOS
and ACS images. Figure \ref{fig:nicmos} shows four examples with
the $i_{775}$ ACS band on the left and the J band from NICMOS on
the right. The impression of a disk dominated by irregular clumps
remains even in the near-IR; there are no surprising central
bulges and the smooth disk component has about the same contrast
to the clumps in both passbands.

These considerations add to the results of Daddi et al. (2005),
Cassata et al. (2005), and Franceschini et al. (2006), who found a
real decrease in spiral and elliptical galaxy co-moving densities
at $z\gtrsim1.5$, even for early spectral types. Considering only
the starbursting types here, Figure \ref{fig:UDF_DANTBDZ} suggests
again that spirals and ellipticals decrease in co-moving density
relatively quickly, but it also shows that intense star formation
in giant clumps causes the irregular types to decrease with $z$
more slowly. In both cases, the spirals and ellipticals that have
smooth surface brightness distributions like modern spirals and
ellipticals are replaced by extremely clumpy galaxies of all
types, including clumpy spirals and ellipticals, at $z\gtrsim2$. A
similar conclusion was reached by Conselice et al. (2005a) who
referred to the clumpy types as peculiar galaxies. We believe it
is important that they are clumpy and not just peculiar in overall
shape.

According to Figure \ref{fig:UDF_DANTBDZ}, the dominant morphology
for $z=1$ to 2 is a clump cluster. At higher $z$, the combined
clumpy types outnumber the combined starburst spiral and
elliptical types by about a factor of 2. The starbursting spirals
dominate only at $z<1$, and then they become only as abundant as
the sum of the clump clusters and chains.  We have previously
noted that spiral galaxies look like smoothed versions of clump
clusters and chains (Elmegreen et al. 2005b), so there should be
an evolutionary effect at work: clump clusters and chains turn
into spiral galaxies if they avoid major mergers.

\subsection{The $V/V_{max}$ Distribution}

The uniformity of a sample of galaxies along a line of sight can
be determined from the distribution function of the ratio of the
survey volume to each object, $V$, divided by the survey volume to
the maximum distance where that object would be included,
$V_{max}$ (Schmidt 1968). If $V/V_{max}$ is uniformly distributed
with an average of around 0.5, then the sample is uniform on the
line of sight. For our sample, $V_{max}$ is given by the redshift
where the object either becomes smaller than our size cutoff of 10
pixels, or becomes fainter than our surface brightness limit,
which is assumed to be 26.5 mag arcsec$^{-2}$ at $z_{850}$ band,
considering that twice the noise rms dispersion is 26 mag
arcsec$^{-2}$.  We find that the surface brightness limit
dominates $V_{max}$.

\begin{figure}\epsscale{0.8}
\plotone{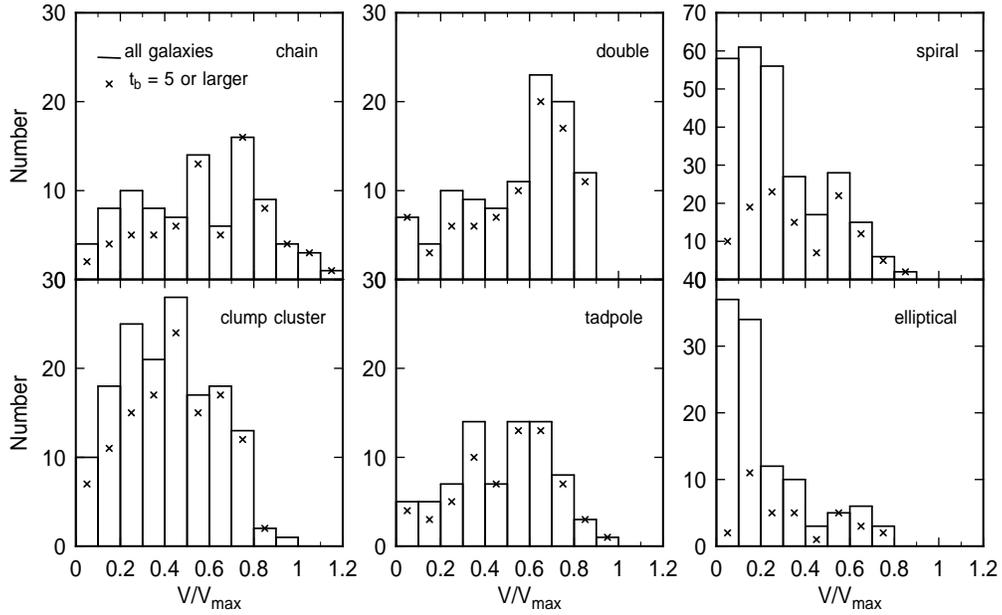} \caption{The distributions of the ratio of the
volume to the observable volume, $V/V_{max}$, are shown for each
morphological type. The solid line histograms are for all galaxies
in our survey and the crosses are for those with starburst spectra
($t_b\geq5$). Considering sampling noise as the square root of the
number of counts, the distribution functions are fairly flat for
all but the spirals and ellipticals, which are relatively flat for
the starbursts and decreasing for all types. The doubles have a
slightly increasing $V/V_{max}$ function, indicating a prevalence at
higher redshifts.} \label{fig:danvvmax}\end{figure}

Figure \ref{fig:danvvmax} shows histograms for the $V/V_{max}$
distributions of all spectral types $t_b$, and shows crosses for
the $V/V_{max}$ distribution of starbursting types, $t_b\geq5$.
For the clumpy types, the two distributions are about the same
because most have $t_b\geq5$; they show that there is an
approximately uniform distribution of these types out to the limit
of the survey, $z\sim5$. For the spirals and ellipticals,
$V/V_{max}$ for all spectral types peaks at low values, indicating
that most of these types are at low redshift, as also seen in
Figure \ref{fig:UDF_danpz}.  For the starbursting spirals and
ellipticals, the distribution is more uniform with a drop finally
at $V/V_{max}>0.8$. These results are consistent with the results
in the previous section which suggested that we are seeing most of
the clumpy galaxies out to the
redshift ($z\sim5$) where the optical restframe is bandshifted beyond
the ACS survey. We are also seeing most of the
starbursting spirals and ellipticals in this survey, except for
those at highest redshift.

\subsection{S\'ersic Index Distributions}
\begin{figure}\epsscale{0.8}
\plotone{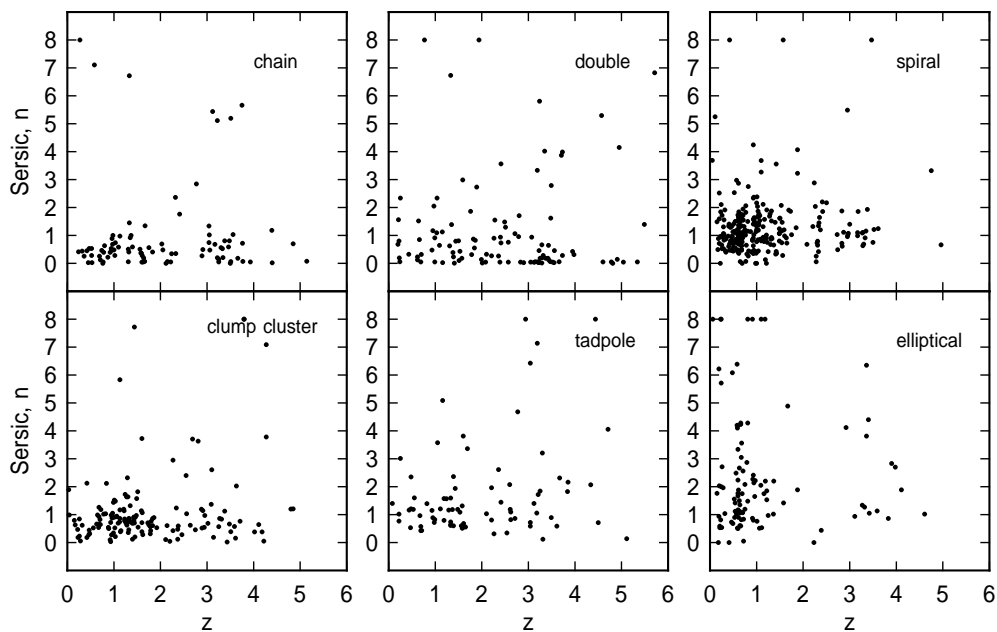} \caption{Distribution of S\'ersic index $n$ as a
function of redshift for different morphological types. The range
of S\'ersic index for ellipticals is larger in the UDF than it is
locally. } \label{fig:UDF_DANNZ6}\end{figure}

Figure \ref{fig:UDF_DANNZ6} shows the redshift distribution of the
S\'ersic index $n$.  Our selection of galaxies with $>10$ px
diameter ensures that they are resolved well enough to give
reliable S\'ersic parameters. Moth \& Elston (2002) suggest that
when the effective radius drops below $\sim2$ px for an elliptical
galaxy, the S\'ersic index drops from $n\sim4$ to $n\sim1$ as a
result of poor resolution. All our galaxies have effective radii
comparable to or larger than 5 px (see Fig.
\ref{fig:UDF_DANSIZE}).

In Figure \ref{fig:UDF_DANNZ6}, the chain, clump cluster, and
double galaxies have a relatively large concentrations with $n <
1$.  These galaxies are dominated by clumps without much variation
in the underlying light (Elmegreen, et al. 2004b), so their
S\'ersic indices are close to zero, which corresponds to a flat
profile inside the half-light radius and a steep profile beyond.
Ravindranath et al. (2006) also found low $n$ for multicore
objects. Cassata et al. (2005) found low $n$ for what they call
perturbed spirals and irregulars/mergers. The tadpole distribution
is similar to the chain and clump cluster distribution, but it
spreads to higher $n$. The high-$n$ tadpoles could have de
Vaucouleurs profiles in the bright clumps that define their
comet-like heads.

The spirals concentrate around $n=1$ with standard exponential
disks. A few spirals with higher $n$ are probably dominated by a
bulge.  The ellipticals have a similar concentration around
$n\sim1$, but also extend to higher $n$, indicating more centrally
concentrated inner radial profiles and flatter outer radial
profiles. There is no particular tendency for ellipticals to have
the de Vaucouleurs profile, $n=4$, even at small $z$. It is
surprising that many galaxies with the appearance of ellipticals
have near-exponential light profiles. A few of these could be
early type structureless disks (S0's), but not the majority of
them because the distribution of axial ratios for galaxies
classified as ellipticals in the UDF is the same as it is locally
(EERS).

Cassata et al. (2005) found that 40\% of what they call normal
spirals with $z<2$ have S\'ersic indices larger than 2, and 10\%
of their ellipticals with $z<2$ have S\'ersic indices less than 2.
Our $n>2$ spiral fraction at $z<2$ is much lower than 40\%. This
is not a selection effect in the classification of spirals because
any galaxy with a bulge, spiral arms, and a decreasing radial
light profile was considered to be a spiral. This would include
$n=4$ disks if there were any with bulges and spirals, but there
are evidently not many of these. Our $n<2$ fraction for
ellipticals is higher than that of Cassata et al., especially at
$z>2$. Hatziminaoglou et al. (2005) found that 30\% of both
spirals and ellipticals are mis-classified by the S\'ersic index
(with $n=2$ as the boundary). They suggested some low-$n$
ellipticals are S0 types and others are too small to get a good
$n$ value; they also reported that large-$n$ spirals have either
dominant bulges or are too small to measure. Our measurements
should not suffer from size effects (see above). di Serego
Alighieri et al. (2005) studied 18 early type galaxies in clusters
at $z\sim1$ and found a wide range of $n$ ($\sim30$\% have
$n\leq2$), but they also pointed out a correlation, present in the
Coma cluster too, where lower luminosity galaxies have lower $n$,
even as low as 1.  A similar $n$-luminosity correlation was noted
by Cross et al. (2004) for E/S0 galaxies from $z=0.5-1$.
Simulations of galaxy mergers that make ellipticals, and
observational studies of mergers, show there should be a wide
range of S\'ersic indices for these types (Aceves, Velazquez, \&
Cruz 2006).

\begin{figure}\epsscale{0.8}
\plotone{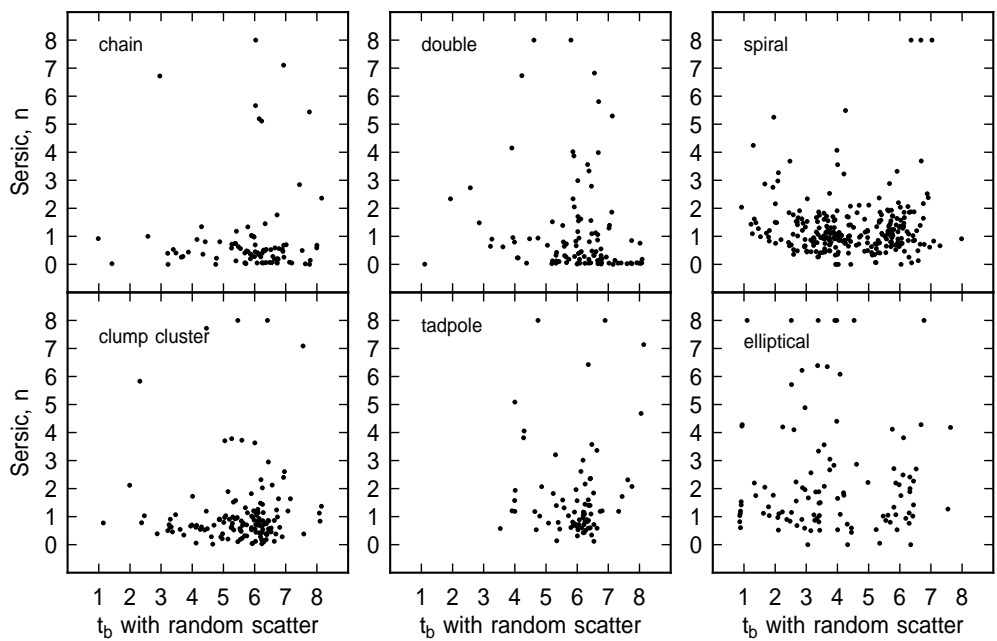} \caption{S\'ersic index versus spectral type (a
random scatter of $\pm0.15$ has been added to $t_b$ to avoid
overlapping points). The radial profiles are not as well
matched to the spectral index in the UDF as they are for local
galaxies. } \label{fig:UDF_danstb6}\end{figure}

The S\'ersic index is plotted versus the spectral type in Figure
\ref{fig:UDF_danstb6}.  For clarity in plotting, the $t_b$ points
are offset randomly around their index values ($\pm0.15$).  There
is wide range of S\'ersic indices for each $t_b$. The indices for
ellipticals range from less than 1 up to 8 in our sample,
independent of spectral type. The spirals typically have
$n=1\pm1$, as do local spirals, also independent of spectral type.
The chain, double, tadpole, and clump cluster galaxies generally
have $n<1$ and $t_b\sim6$, as noted above. Evidently, spectral
energy distributions of resolved galaxies in the UDF are not
uniquely linked to their morphology, and neither property
correlates well with the form of the radial light profile. The
reason for this is that ellipticals are not well distinguished by
their S\'ersic profiles, and both modern types, spirals and
ellipticals, have a wide range of star formation rates.

\subsection{Size Evolution}

Galaxy growth over redshift is expected to be significant in the context of hierarchical
galaxy formation, although current observations show that for $z<1$ the size remains fairly
constant.  Ravindranath et al. (2004) found that
disk galaxies in the GOODS-S field at $0.25<z<1.0$ do not show
significant size evolution when selection effects are considered.
Barden et al. (2005) got a similar result for $z<1.1$ disk
galaxies in GEMS and concluded that size has to increase with mass
to keep the mass-radius relation constant over redshift. Trujillo
\& Pohlen (2005) observed a 25\% growth in the outer disk
truncation radius of UDF spirals over this redshift range, while
Pirzkal et al. (2006) found essentially no evolution in size from
$z=1.5$ to 0 for small, blue UDF galaxies. Stockton et al.
(2006), on the other hand, found two red spiral galaxies at
$z=1.4$ that are about half the size of local spirals.

Many studies have shown that a factor of 2 in growth is typical for longer redshift intervals.
Roche et al. (1998) found that size evolution begins to appear at
$z>1.5$ but primarily for the star-forming galaxies. Cassata et
al. (2005) observed the opposite: ellipticals are a factor of
$\sim2$ smaller at $z=2$ than $z=0.5$ but irregulars have a
constant size. Daddi et al. (2005) obtained the same growth factor
since $z\sim1.4-2.5$ in his sample of 7 early type galaxies, one
of which could be a spiral. Trujillo et al. (2006a) examined
galaxies at $z < 3$ in the SDSS, GEMS, and FIRES surveys and found
that S\'ersic $n<2$ galaxies are smaller than they are today for a
given luminosity by more than a factor of 2 at $z=2.5$; $n>2$
galaxies were nearly 3 times smaller at $z=2.5$ than equal
luminosity galaxies today. Trujillo et al. (2006a) got a factor of
$\sim2$ variation for a given mass since $z\sim2.5$ for both low
and high-$n$ galaxies. Other evidence for $2\times$ growth from
$z=4$ to 1.5 was seen in a sample of galaxies of all types from
the HDF-N and CDF-S fields (Ferguson et al. 2004) and for a sample
of disk galaxies from the HDF-S (Tamm \& Tenjes 2006). Bouwens et
al. (2003) found a 1.7 times increase in size for a mixture of
morphological types from $z=5$ to $z=2.7$ based on HDF data, and
Bouwens et al. (2004a; 2006) observed size evolution from $z=6$ to
$z=2.6$ for a mixture of types in the UDF data. Conselice et al.
(2003) also found a factor of 2 variation in size for the largest
galaxies with a mixture of morphologies in the HDF between $z=1$
and 4.

Some of this size evolution could be the result of restframe
surface brightness evolution because galaxies at high redshift are
starbursting for all $n$ (see above). This makes their surface
brightnesses high, and then intrinsically small galaxies can have
the same luminosities as present-day large galaxies, even if there is
no growth over time. Some of the apparent size evolution could be
from size-of-sample effects considering that the size distribution
function favors small galaxies.  Then a limited sample of galaxies
would have a smaller average size and a smaller largest galaxy
size than a big sample of galaxies.

We are interested in the size evolution for each morphological
type. We previously measured the distribution of exponential scale
lengths for all resolved spiral galaxies in the UDF, without
separation into redshift intervals, and found the average length
to be about half that for local spirals (Elmegreen et al. 2005b);
this is consistent with the other
results mentioned above because spirals concentrate at
$z\sim1$ (Fig. \ref{fig:UDF_danpz}). Here we consider
galaxy size as a function of both morphological type and redshift.
The clumpy morphological types require special definitions of size
because S\'ersic fits and the associated effective radii may not
be appropriate for such irregular structures.  We also want sizes
that are independent of surface brightness dimming. For spiral and
elliptical galaxies, this means the total light profile has to be
extrapolated to infinity, beyond the visible edge, in order to
find the half radius of the total, extrapolated light. This is the
standard procedure in GALFIT, which was used here. However, the
light profile is neither regular nor predictable beyond the last
clump in a highly clumpy galaxy, so again, $r_{eff}$ is not a good
measure for these types.

\begin{figure}\epsscale{0.8}
\plotone{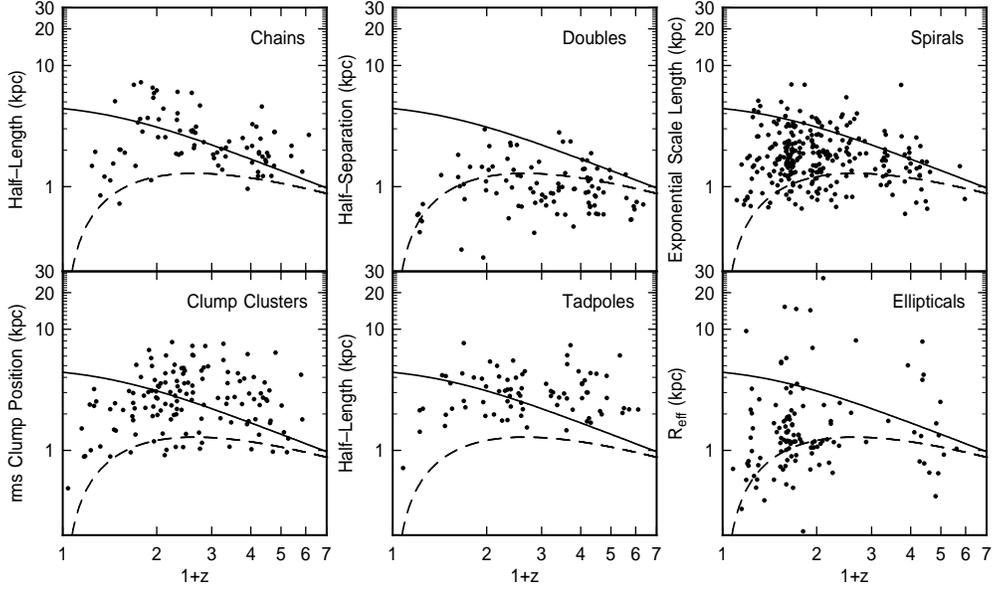} \caption{Radius versus $1+z$ for the 6
morphological classes. Radii are half-light effective radii for
ellipticals, exponential disk scale lengths for the spirals, and
half of the end-to-end dimensions for tadpoles, doubles, chains,
and clump clusters. The solid curve is the size of a $10^{10}$
M$_\odot$ galaxy at 200 times the average density of the Universe,
from Mo, Mau \& White (1998). The dashed curve is the size of 5
pixels, the minimum 2-$\sigma$ major-axis contour radius of our
survey.  The mean sizes are about constant with redshift, although
the maximum sizes tend to decrease slightly to higher $z$,
especially for the chains and spirals.}
\label{fig:UDF_DANSIZE}\end{figure}

\begin{figure}\epsscale{0.8}
\plotone{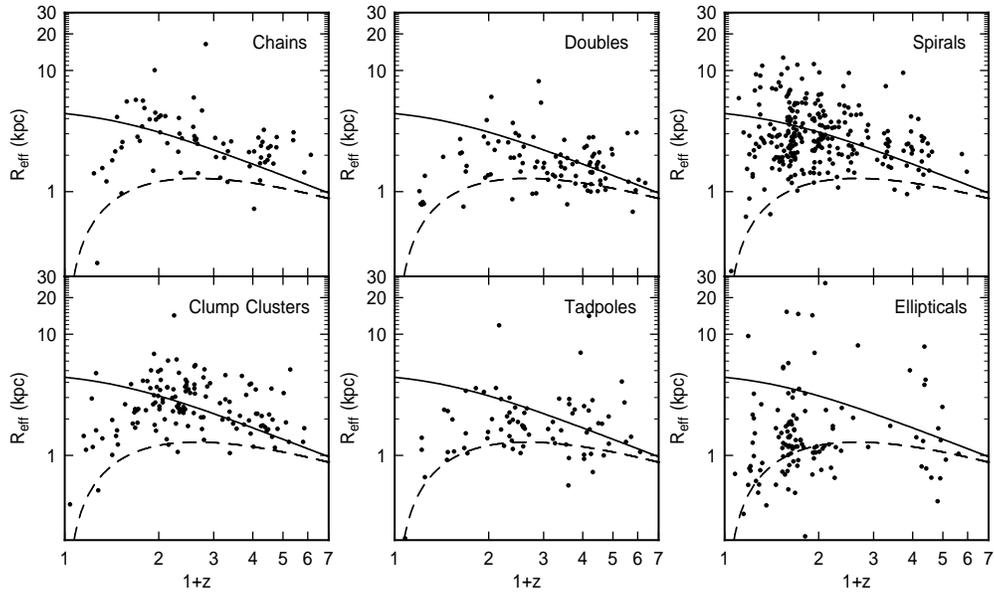} \caption{Effective radii $r_{eff}$ determined
from GALFIT light profiles are shown versus $1+z$ for the 6
morphological classes. For doubles and other highly clumped
galaxies, $r_{eff}$ measures the galaxy size only crudely. The
results are essentially the same as in Figure
\ref{fig:UDF_DANSIZE}.} \label{fig:UDF_danreff}\end{figure}

Figures \ref{fig:UDF_DANSIZE} and \ref{fig:UDF_danreff} show size
distributions versus $1+z$ for each type, with sizes measured
differently in the two figures.  The solid curve in each is the
size of a $10^{10}$ M$_\odot$ galaxy at 200 times the average
density of the Universe, from Mo, Mau \& White (1998). The dashed
curve is the size of 5 pixels, the minimum 2-$\sigma$ major-axis
contour radius for our survey.

In Figure \ref{fig:UDF_DANSIZE}, the sizes for the chains,
tadpoles, and doubles are defined to be half the separations
between the main clumps at the ends of the linear structures. For
clump clusters, the sizes are defined to be the deprojected rms
separations between the clumps (deprojection assumes circular
galaxies and uses the measured major and minor axis lengths). The elliptical sizes were measured in the usual way,
using the half-light effective radii $r_{eff}$ defined by GALFIT.
The spiral sizes were measured in a somewhat standard way, using
their exponential disk scale lengths (from Elmegreen et al.
2005b). Because these sizes are measured in different ways for the
different morphologies, the distributions should not be directly
compared to each other in Figure \ref{fig:UDF_DANSIZE}.

In Figure \ref{fig:UDF_danreff}, the sizes are taken to be
$r_{eff}$ from GALFIT. Even in this case, comparisons between the
different morphologies should be done with care. For example,
$r_{eff}$ for a double galaxy, each part of which looks
spheroidal, is proportional to the separation between the clumps
and nearly independent of the size of each one.

The relative size distributions are about the same in both
figures. The chains, spirals, and clump clusters show a decrease
in size by about a factor of 3 from $z=1$ to 4, roughly following
the solid line. The doubles and ellipticals show a factor of 2
size decrease from $z=2$ to 5. The tadpoles have widely scattered
size distributions with no obvious trend. This lack of evolution
for tadpoles may not be surprising if tadpoles are interacting
galaxies with long tidal tails; the sizes we measure are
essentially the lengths of the tails. For all types, the lowest
redshift galaxies ($z<0.25$) are among the smallest, but this is
probably a selection effect for the choice of UDF field (which
avoided galaxies that are large in angular size).

The average size distributions for galaxies cannot be determined
from these figures because of the constant lower limit from our
sample criterion.  The trends in the upper limits are not good
indications either because of possible size-of-sample effects.
That is, there are more low-$z$ galaxies than high-$z$ galaxies,
and for the galaxy size distribution function, which has
more smaller galaxies than larger galaxies,
the size of the largest galaxy in a big sample (the low-z sample) is likely to be
larger than the size of the largest galaxy in a small sample (the high-z sample).
This could explain our observed size-redshift trend
even if the intrinsic size distribution at each redshift is
constant. Still, the trends in the figure suggest that, aside from
the tadpoles, the observed UDF galaxies increase in size by
factors of 2 to 3 from $z\sim5$ to $z\sim1$.

\section{Discussion}
\label{sect:disc}

The redshift distributions of the spectral types, S\'ersic indices,
and sizes of galaxies in the ACS image of the UDF have been shown
for 6 galaxy types, which are the usual spiral and elliptical
types plus four peculiar types that are characterized by their
clumpy structures. The UDF spirals and ellipticals have a
relatively small range of redshifts, centered on $\sim0.2-1.6$,
while the clumpy galaxies have a broad range, from $\sim0.2$ to 5.
Most of the clumpy types have starburst spectra, and all of the
$z>2$ spirals and ellipticals in our sample are starbursts.

Starburst galaxies in the UDF in every redshift interval are dominated by
the four clumpy types, which are the chains, clump clusters, tadpoles and doubles. The comoving density (Fig.
\ref{fig:UDF_DANTBDZ}) of $t_b\geq5$ clump clusters and chains
exceeds the comoving density of $t_b\geq5$ spirals and ellipticals
in every redshift bin beyond $z=1$ by an average factor of $1.8$.
The comoving density of all $t_b\geq5$ clumpy types exceeds that
of $t_b\geq5$ spirals and ellipticals by a factor of 3.2 for
$z>1$. For $z<1$, $t_b\geq5$ spirals dominate all other starburst
types, but the combination of the $t_b\geq5$ clumpy types exceeds
the combined $t_b\geq5$ spirals and ellipticals by a factor of
1.2.

These observations indicate that star formation occurs in giant
clumps in the disk-like UDF galaxies that have no bulges or exponential
light profiles. These clumpy disks span the full range of
observable redshifts, out to at least $z\sim5$, and presumably
form, merge, and evolve into spirals and ellipticals continuously
over this time. Considering the sizes and separations of the
clumps, the most likely mechanism for clump formation is a
gravitational instability in the gaseous component of the disk or
in the shock between interacting disks. This scenario follows
from our previous observations (EE05, EE06), which showed that
clumps have sizes and masses consistent with the Jeans length and
mass in a gas layer with a turbulent speed of $\sim15$ km s$^{-1}$
or more. For dominant unstable wavenumber $k_J=\pi G\Sigma/a^2$ in
a disk with mass column density $\Sigma$ and velocity dispersion
$a$, the Jeans mass is $M_J=\Sigma \left(\pi/k_J\right)^2\sim
a^4/G^2\Sigma$. Taking $\Sigma\sim15$ M$_\odot$ pc$^{-2}$, as
observed for chains and edge-on spirals in the UDF and for modern
thick disks, and $a\sim15$ km s$^{-1}$ to explain the disk
thicknesses at high redshift (see EE06 for both measurements), we
get $M_J\sim2\times10^8$ M$_\odot$, in good agreement with the
observed clump masses (EE05).

Clump masses in clump cluster and chain galaxies are a factor of
$\sim100$ larger than cluster and star complex masses in modern
disk galaxies, indicating that star formation was more violent at
high redshift than it is today. The clumpiest disks also extend to
higher redshifts than the smooth disks, so the clumpy disks are
probably the most primitive. Most likely, the clumps in clumpy
disks dispersed and mixed to make the smooth disks (see also
Elmegreen et al. 2005b). This would presumably happen even at the
highest redshifts in our survey, $z\sim5$, thereby accounting for
the giant red spirals at $z\sim2$, and following mergers, the giant
red ellipticals as well. This scenario is consistent with the
stellar evolution models by Fritz von Alvensleben \& Bicker (2006)
and the observations by Bundy et al. (2005, 2006) where spirals
turn into ellipticals after ending their star formation some 1-2
Gyr earlier. We note that during mergers, star formation can occur
in spheroidal geometries if the original disk components warp and
scatter their giant clouds into new orbits that fill the volume of the spheroid.
Still, the main star formation mechanism is likely to be one
preferential to disks because that is where the dense gas is.

The constancy of the comoving density and the $V/V_{max}$ ratio
for tadpoles and doubles suggests that galaxy interactions and
mergers were common back to $z\sim5$. The tadpoles are presumably
interacting galaxies with long tidal tails (e.g., Straughn et al.
2006) and the doubles are presumably near-neighbor spheroidals in
the process of merging (Tran et al. 2005; van Dokkum 2005; Mei et
al. 2006; Bell et al. 2006).

Clump cluster and chain galaxies are only moderately luminous and
should therefore not cluster as strongly as the massive galaxies
of this epoch (Sect. \ref{sect:mag}). According to cosmological
simulations, galaxy brightness at high redshift correlates with
dark matter mass, and this mass increases as the the scale length
increases for 2-point correlations with other bright galaxies. For
this reason, bright galaxies at high redshift should end up as
massive ellipticals in rich clusters where interactions have been
strong (see Adelberger et al. 2005). Without strong clustering,
the clump cluster and chain galaxies at low and intermediate
redshifts could evolve in relative isolation, possibly forming
modern spirals after continued growth through accretion and minor
mergers. They would seem to make only late-type spirals, however,
because they have no bulges. Bulges could result from clump
coalescence (Noguchi 1999; Immeli et al. 2004a,b), or bar
dissolution (Hasan \& Norman 1990), but then there is no clear way
to make nuclear black holes unless they are already inside the
disk clumps and merge or they grow in the nuclei over extended
periods of time from accreted disk gas.  The giant clumps do not
appear dense enough to make black holes, however (e.g., see
Ebisuzaki et al. 2001). Their average densities correspond to only
a few tens of atoms cm$^{-3}$ (EE05). The second possibility, slow
growth of black holes, would not obviously account for the
bulge-black hole mass correlation. If, having no bulges, clump
clusters and chains form only late-type spirals, then where are
the predecessors of early-type spirals at $z\sim5$? Early-type
spiral progenitors would seem to be our clumpy disks at the
highest redshifts, because these would convert their gas into
stars and fade relatively early. They are still lacking bulges,
though.

It is conceivable that most of the clump cluster and chain
galaxies at $z\ge1$ in the UDF do not survive in relative
isolation, dissolving and blending their clumps over time to make
smooth exponential disks. Instead, they could undergo major
mergers that either form spheroids or, in the case of in-plane
orbital motions, force the gas and stars into the nuclear regions
to make black holes and bulges surrounded by a residual
disk (e.g., Hopkins et al. 2006). If this is the case, then these
peculiar types are temporary states that give us a snapshot of
what star formation in isolated galaxies was like in the early
Universe.

\section{Conclusions}

Spectral energy distributions, $t_b$, S\'ersic indices, galaxy
morphologies, and galaxy sizes were studied as functions of
redshift for galaxies in the UDF. Galaxies that appear to be
elliptical have the full range of S\'ersic indices with a
concentration around $n=1.5\pm1.5$ for all $z$. The ellipticals
also have a wide range of spectral types at low $z$. At high $z$,
all observed galaxy types have starburst spectra because our selection
in ACS optical bands corresponds to the restframe UV wavelength.
The spiral galaxies have S\'ersic
indices around $n=1\pm1$ and a
redshift distribution of spectral types similar to that of
ellipticals: a broad range at low $z$ and only starbursts at high
$z$.  The comoving volume density of starbursting spirals and
ellipticals falls off rapidly with redshift. The fall-off is
slower for the clumpy morphologies, which are generally
starbursting and which dominate every redshift interval in the
UDF. The $V/V_{max}$ distribution shows the same trends. Red
spiral and elliptical galaxies at $z>2$ are not observed in the
ACS because of bandshifting. The sizes of all galaxies except for
tadpoles increase by a factor of $\sim2$ from $z=4$ to 1.
Size-of-sample effects and the minimum size cutoff prevent
definite conclusions about galaxy growth rates at this time.

The observations here and in our previous papers (EE05, EE06)
support a model in which star formation occurs primarily in disks
by gravitational instabilities having a Jeans mass of
$\sim10^7-10^9$ M$_\odot$ and a Jeans length of several kpc. The
disks extend back to at least $z\sim5$. Clump clusters and
chains are examples of star-forming disks at their earliest stage.
The comoving density of these two types combined exceeds that
of starbursting spirals and ellipticals at $z>1$ by a factor of 1.8.
Clump clusters and chains appear to be the initial conditions for
spiral galaxies and ultimately, through successive mergers, for
spheroids as well. There is no evidence for a star formation mode
like what is usually assumed for isolated monolithic collapse;
i.e., dispersed and spheroidal primary sites with a gradual
settling of the associated clouds to a disk. Instead, star
formation appears to begin in disks at high redshifts and then
become spheroidal later, if it does at all, by three-dimensional
scattering and shocking of the disk clouds during major mergers.

The lack of bulges in clump cluster and chain galaxies is a
mystery. Either these types form bulges late, after the primitive
disk stage we see in the UDF but before they become modern
spirals, or they get significantly reorganized by mergers into
ellipticals and spiral galaxies with bulges.

\acknowledgements D.M.E. thanks the staff at Space Telescope
Science Institute for their hospitality during her stay as a
Caroline Herschel Visitor in October 2005. D.M.E. also thanks
Vassar for research support through the Olin fund. Helpful
comments by the referee are gratefully acknowledged.

\newpage

\end{document}